# Look What You Made Me Glue – SEMGlu™ Enabled Alternative Cryogenic Sample Preparation Process for Cryogenic Atom Probe Tomography Studies

Neil Mulcahy, James O Douglas,* Michele Conroy*

Department of Materials, London Centre of Nanotechnology, Imperial Henry Royce Institute, Royal School of Mines, Imperial College London, SW7 2AZ, UK

*Corresponding author email: j.douglas.imperial.ac.uk, m.conroy@imperial.ac.uk

## Abstract

Extensive efforts over the past number of years have been applied to develop workflows for sample preparation of specimens for atom probe tomography at cryogenic temperatures. This is primarily due to the difficulty involved in preparing site specific lift out samples at cryogenic temperatures without the assistance of the gas injection system (GIS) as using it under cryogenic conditions leads to non-uniform and difficult to control deposition. Building on the efforts of previously developed GIS free workflows utilising redeposition techniques, this work provides an alternative approach using SEMGlu™, which is an electron beam curing adhesive that remains usable at cryogenic temperatures, to both lift out cryogenically frozen samples, and mount these samples to Si microarray posts for subsequent redeposition welding. This approach is applicable for a full cryogenic workflow but is particularly useful for non-fully cryogenic workflows such as beam sensitive samples, samples that mill easily, and samples with challenging geometries. We demonstrate atom probe analysis of silicon samples in both laser pulsing and voltage mode prepared using this workflow, with comparable analytical performance to a pre-sharpened microtip coupon. An application-based example which directly benefits from this approach, correlative Liquid Cell Transmission Electron Microscopy and cryogenic Atom Probe Tomography sample preparation, is also shown.

## Introduction

The emerging field of atom probe tomography (APT) in combination with cryogenic sample preparation and cryogenic/inert gas/vacuum transfer has gained increased popularity in recent years due to atom probes unique capability of providing three-dimensional compositional mappings of frozen nanoscale volumes with high-spatial and high-compositional resolution[1-4]. This has proven particularly useful for investigating a growing number of material systems including various battery materials [4, 5], hydrogen embrittlement in steel and other metals [6, 7], as well as various frozen liquids and frozen liquid-solid interfaces [8-11].

Site specific sample preparation for APT is typically carried out using Focused Ion Beam (FIB) lift out and sharpening, with the use of a decomposed organometallic gas via a Gas Injection System (GIS) acting as a conductive "glue" [12, 13] to both lift out and mount samples. Cryogenic sample preparation for APT studies has additional challenges and remains an area in active development. At cryogenic temperatures, the GIS will produce uncontrolled condensation of frozen precursor species, without decomposition onto the substrate [13]. There is some possibility of control as detailed by Parmenters et al. [14] but more typically this uncontrolled deposition will condensate over a mm$^2$ size area of the surface of the cold sample, losing the site specificity of the process, while simultaneously depositing a thick layer of organic and metal-organic species [15, 16]. This lack of site specificity makes it extremely challenging to identify a site-specific region of interest (ROI) and upon lifting out easily attaching this ROI to a Si microarray post at cryogenic temperatures. While significant progress has been made in achieving a GIS free lift out procedure using selective redeposition of a conductive material [3, 17], which has now been applied to a number of environmentally sensitive samples including free standing frozen water [3], several problems still remain. The process is not as easily translatable to materials that are beam sensitive, mill extremely easily with respect to the support structure, or those with unique geometries that cannot achieve a defined undercut that can be filled with sputtered material. Alternative GIS-free methods taking advantage of the fast-milling rates and high beam currents of $Xe^+$ plasma FIB instruments have been demonstrated and avoid the lift out procedure entirely through annular milling of trenches directly into a flat sample to leave a free-standing needle shape appropriate for AP analysis [18]. This "satellite" method has been used to investigate various materials systems and has been shown to be a viable site-specific sample preparation method at cryogenic temperatures [10]. However, this methodology is only applicable to certain material systems most notably nanoporous systems, and has proven challenging for free standing, thin and liquid based samples. This methodology is also only viable with the use of a PFIB, as using a $Ga^+$ FIB will require long milling times at exceptionally high currents, leading to large amounts of $Ga^+$ implantation [18].

Avoiding the use of the GIS, or indeed not having a GIS or micromanipulator within the FIB system is not a new concept in FIB site specific lift out sample preparation and a number of solutions have been developed over the past number of decades for a number of applications. Ex-situ Lift Outs (EXLO) for preparation of STEM/TEM lamellae have been standard practice for many different material system applications for a number of years [19, 20]. This method is performed outside the vacuum of the FIB chamber, through the use of an optical microscope and relies upon adhesion forces between the prepared lamella and typically a glass needle to lift the lamella out and place it on a conventional TEM grid or onto a slotted grid carrier [19]. It has been found that at length scales <100 μm adhesion forces are dominated by Van der Waals and Capillary forces for conducting samples [21]. At these length scales, the force of adhesion is greater than that of gravity, meaning lamellae can be easily manipulated. Insulating samples are more difficult to manipulate using this method due to electrostatic forces causing samples to either be repulsed by the glass needle or attracted to the bulk sample. This has been combatted by using a metal tip or coating the glass needle with an appropriate metal [22]. EXLO is not limited to specific instruments and has been applied using lamella prepared in both Ga FIBs and Xe PFIBs [19, 23]. Further to this Cryo EXLO has also been demonstrated [24].

EXLO has been applied to a number of materials applications but has proven to be particularly useful when applied to placing prepared lamellae on MEMs based chip devices for use in in-situ TEM applications such as in-situ electrical biasing [19]. E-beam/FIB-beam induced deposition has been shown to inadvertently allow conductive material to spread across these chip devices, ultimately resulting in the development of stray leakage current pathways, hindering representative electrical performance [25]. EXLO avoids this by placing the lamella directly on the contact/window/ROI. Adhesion using locally decomposed carbon from the chamber has been shown to be a viable method for sticking down the lamella or alternatively purely through Van der Waals [26, 27]. Other innovative methods allowing for the use of the GIS to lift out the sample and attach it to the ROI on the MEMs chip using adhesive forces or Pt have also been developed [28, 29].

The use of carbon deposition via the e-beam as against via the GIS with FIB deposition has also found use in many other applications, particularly with respect to mounting TEM lamellae [30, 31]. Using e-beam deposited carbon has been shown to reduce beam damage next to the region of the weld section. However, this still leaves a layer of adsorbed gas present on exposed areas. If this is then exposed to any secondary electrons from either an electron or ion beam, this can lead to unintended deposition of a thin layer of material [31]. E-beam deposited carbon can also be used to attach carbon nanotubes or other extremely small samples to manipulators [32]. However, it is unlikely to be appropriate for lifting out entire lamellae. It has been shown that it is also possible to deposit entire samples directly onto conventional TEM grids and to directly thin these deposited samples to electron transparency. This method avoids having to produce a lift out site-specific section, meaning the use of the GIS can also be avoided. This is referred to as "On-grid thinning" and has received widespread use in biological fields, particularly for producing lamella at cryogenic temperatures to perform cryo TEM and cryo-electron tomography [33]. However, this methodology requires extensive specialised instrumentation that may not be transferrable to material science applications.

Alternative approaches ex-situ have been developed for the preparation of samples for conducting APT. Electrophoresis has been used to generate APT samples of nanoparticles [34]. For this a metal tip is placed in a dispersion of nanoparticles, while a large negative voltage is applied. The nanoparticles are attracted to the metal tip due to the generated electric field and preferentially attach to the apex of the tip. It has been shown that this method will generate a sufficient bond between the tip and the particle such that no additional sample preparation is required, including coating [34]. This is often only applicable to really small samples, with larger samples often requiring some degree of coating to withstand detachment while an electrostatic stress is applied during AP analysis [34]. Alternative methods for larger particles include picking up and placing nanoparticles using an SEM and a micromanipulator on a sharp tip, followed by PVD deposition to encapsulate the particle of interest and eventually sharpening using a FIB [35, 36]. The concept of encapsulating samples in order to withstand AP analysis and subsequently generate useful data has been extended to other sample preparation workflows, both ex-situ and in-situ at room temperature, through the use of processes such as Atomic Layer Deposition (ALD) and in-situ sputtering from a conductive material. This has allowed for various nanoparticles [37], 2D materials [38] and battery materials [39] to be investigated using APT, which was previously not possible/difficult.

Of course, the above examples are for specific applications and are adequate for the desired length scales, samples, and conditions. Many of the above methodologies for FIB lift out and mounting would not meet the required mechanical, thermal, and electrical requirements needed to withstand the large applied electrostatic stress during APT data acquisition and acquire sufficient data quality for a site-specific lift out sample [40], particularly a cryogenically frozen ROI. To that end, more robust (sticky) methodologies were needed to meet these requirements. Several adhesive based solutions have been demonstrated to achieve this. Adhesives are appropriate for nanoscale manipulators and have been used for both ex-situ and in-situ applications [41, 42]. Key aspects for consideration when selecting an appropriate adhesive include vacuum stability, curing time, conditions, and sample of interest. An early example of the use of an adhesive to prepare superconducting oxide samples for "FIM-FEEM" was by A.Melmed [43]. Crushed sharp fragments were picked up using a manipulator coated in a conductive adhesive without further sample preparation. This work was adapted for modern APT sample preparation by D. Larson et al. [44], which involved depositing thin film structures on Si wafer posts, fabricating posts, fracturing these posts and manipulating them through the use of silver containing epoxy on a metal manipulator with an optical microscope setup, similar to the EXLO described previously. The picked-up sample is subsequently straightened and placed in a FIB for sharpening prior to atom probe analysis.

Various adhesives have been used in the following years for both TEM and APT sample preparation such as SemGlu$^{TM}$ and cyanoracrylates (superglue). SemGlu$^{TM}$ is a vacuum safe adhesive which cures into a conductive solid under electron beam irradiation [30, 45, 46] and has been used as a glue for room temperature atom probe sample preparation with success [47]. It does have a few caveats with respect to evidence of increased thermal tail artefacts reported in short (under 3 μm) lift outs and this was linked to the decreased thermal conductivity of the SemGlu$^{TM}$ after curing [47]. Increased carbon contamination has also been reported, especially if there is significant uncured SEM glue left on the sample prior to analysis [47]. While SEMGlu$^{TM}$ isn't necessarily inherently designed for use under cryogenic conditions, typical "cryogenic" adhesives are often required to be cured under non cryogenic conditions anyway [48]. However, low temperature and cryogenic adhesives do exist and have found widespread use in aerospace, space, refrigeration, and freezer applications [49, 50]. The first use, to the authors knowledge, of specifically a cryogenic adhesive for low temperature SEM was by Okada &

Chen [48]. However, no examples for APT/TEM sample preparation using cryogenic adhesives currently exist.

It is common practice for samples to be mounted using a liquid adhesive at room temperature prior to cooling to cryogenic temperatures [51, 52]. At cryogenic temperatures liquid adhesives will freeze but will still act as an adhesive and maintain the orientation of the sample of interest. The use of liquid adhesives under cryogenic conditions can of course lead to a number of issues, particularly with respect to a reduction in viscosity with decreasing temperatures [48]. The exact temperature used within a cryogenic FIB stage is limited by the refrigerant, with liquid nitrogen generally being the refrigerant of choice, due to its convenience, cost, and availability. The selection of the base temperature for the system is dependent on the material system being studied and so the stage and manipulator may not need to be at the lowest temperature possible from the system. With the use of in-situ temperature measurement and control via heaters, certain systems can be tailored to operate just below the freezing point of the liquid adhesive, provided the system of interest remains in a frozen state.

The two main yield limiting stages for the full cryogenic lift out workflow for APT sample preparation are the initial lift out process and mounting of the lift out bar onto a Si microarray post. The work presented here provides alternative solutions to overcome these yield limiting steps using SEMGlu$^{TM}$ at liquid nitrogen temperatures. It has been demonstrated here that pre-coating pre-prepared Si posts with SEMGlu$^{TM}$ using the micromanipulator at room temperature, cooling to liquid nitrogen temperatures and performing both a lift out and mounting process produces reproducible success, as even after the glue has frozen at cryogenic temperatures, it still provides local curing and adhesion. This process flow was applied to both a model Si system, as well as for a specific application-based system involving performing correlative Liquid Cell Transmission Electron Microscopy (LCTEM) and Cryogenic APT. Both are detailed, outlining the key steps in achieving successful SEMGlu$^{TM}$ enabled cryogenic APT sample preparation even for the (un)stickiest of situations.

## Instrumentation and Materials

### FIB with Cryo-Stage

All FIB/SEM work was completed using a Helios Hydra CX (5CX) plasma FIB from Thermo Fisher Scientific (TFS) fitted with an Aquilos cryo-stage. A "Dual-puck" holder stage baseplate supplied by Oxford Atomic was fitted to the cryo-stage allowing for two industry standard cryo pucks supplied by CAMECA Inc. to be mounted directly to the stage, negating the need to vent the SEM chamber or the need to transfer one cryogenic sample out in place of another cryogenic sample. The system is equipped with an EasyLift tungsten cryo-micromanipulator. Both the stage and micromanipulator can be cooled to approximately 90 K through the circulation of gaseous nitrogen passing through a heat exchanger within a dewar of liquid nitrogen (LN2). The flow of gaseous nitrogen was maintained at 180 mg/s to achieve the base temperature for the cryogenic work presented. The temperature of either the stage or micromanipulator could be adjusted using a temperature control unit, Model 335 cryogenic temperature controller supplied by LakeShore, in conjunction with heaters within the stage. Xe plasma was used throughout the work presented. The FIB column was set at 52° to the electron column. The system is also equipped with a Ferroloader docking station. This enables transfer between the SEM chamber and an ultra-high vacuum cryo transfer module (VCTM), that may also be cooled to cryogenic temperatures via a Dewar of LN2 [53].

### Atom Probe

Atom probe analysis was completed using a CAMECA Local Electrode Atom Probe 5000 XR. This system is equipped with a reflectron and a Ferroloader docking station, allowing samples to be directly inserted into the load lock chamber without breaking vacuum. Cryogenic samples may be inserted to the analysis chamber from the VCTM using a cooled "piggy-back" puck maintaining the temperature of the sample. Samples were run using both laser pulsing and high-voltage (HV) pulsing. Data analysis and reconstructions of 3D atom maps were completed using AP suite 6.2.1, a commercially available software provided by CAMECA Inc.

## Materials

SEM-compatible glue or SEMGlu[TM] used throughout this work was supplied by Kleindiek Nanotechnik GmbH [45, 46]. A microarray post, commercially available from CAMECA Instruments Inc., was used as both the substrate for the model Si system discussed and for the mounting of APT samples. These microarrays are made of highly Sb-doped single-crystal silicon and act as a reference for the calibration of CAMECA's commercially available instruments due to the materials high level of conductivity and mechanical stability [54].

The lift-out approach developed here was applied to correlating liquid cell in-situ electrochemistry TEM or x-ray beamline experiments with cryogenic atom probe tomography to investigate dendrite growth mechanisms, morphology, and compositions in lithium-based battery systems. For the purpose of this work, the liquid cell chips were acquired commercially from DENSsolutions and create a "Nanobattey" with a series of platinum working electrodes on a thin silicon nitride membrane within a silicon wafer chip. For this work, the in-situ LCTEM electrochemistry experiment was conducted using a commercially acquired lithium electrolyte, in this instance $LiPF_6$ in Propylene carbonate (PC), supplied by Merck Life Science UK Ltd. By cycling the applied voltage, dendrites grew from the Pt electrodes. After 3 cycles the experiment was stopped when large dendrites formed. The beam was then blanked and the in-situ holder was left in the TEM for 30 min to ensure the dendrites did not dissolve over time. The holder was transferred to an $N_2$ glove box and disassembled where the chip was frozen inside on a cold block and then submerged into a liquid nitrogen bath. The chip was then transferred to the cryo stage of the SEM/FIB using the Ferrovac VCTM set-up described in methods above.

## Results and Discussion

### Pre-preparation and Pre-coating

The entire basis of this lift out and mounting process requires both initial preparation and coating of the Si microarray posts with glue. The process was initially conducted at room temperature due to the viscosity of the SEMGlu[TM]. A series of flat-topped Si posts are prepared to fit a sample that would be undercut at 52° with respect to the Ion beam, first described in [8]. The post is milled at a 0° stage tilt with a 30kV 4nA Ion beam probe, while leaving a section of the post on the right-hand side to act as a support. This is done to increase the area of contact between the post and the lifted-out sample. This extra contact area was deemed necessary due to a large volume of samples being lost when being milled free from the micromanipulator once in contact with the post while performing a GIS free lift out using redeposition techniques. This low mounting yield could most likely be attributed to the reduced resolution of a plasma FIB when performing redeposition [55]. An example of a precut post can be seen in Figure 1.

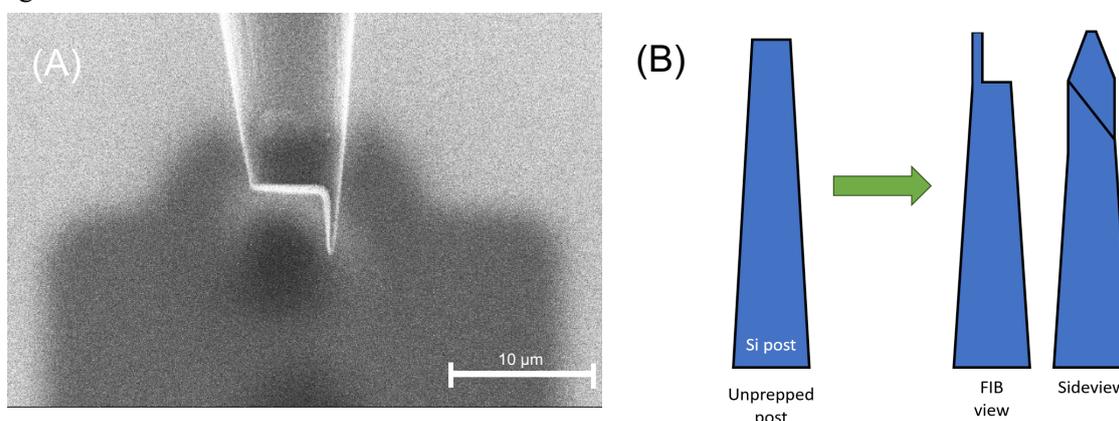

**Fig. 1:** (A) An SEM micrograph of a precut of a precut Si microarray post at 0°, with left side of post untouched to act as support. (B) Diagram with view of geometry of pre-cutting procedure.

The post is further prepared to accept the lift-out by coating it with SEMGlu™. The micromanipulator is inserted into the SEMGlu™ located on the clip holding the microarray post. By moving the manipulator through the SEMGlu™ a large amount will attach itself and can be subsequently "wiped" onto the silicon posts allowing for a small amount of the glue to remain on the posts. This wiping procedure is highlighted in Figure 2(A-F).

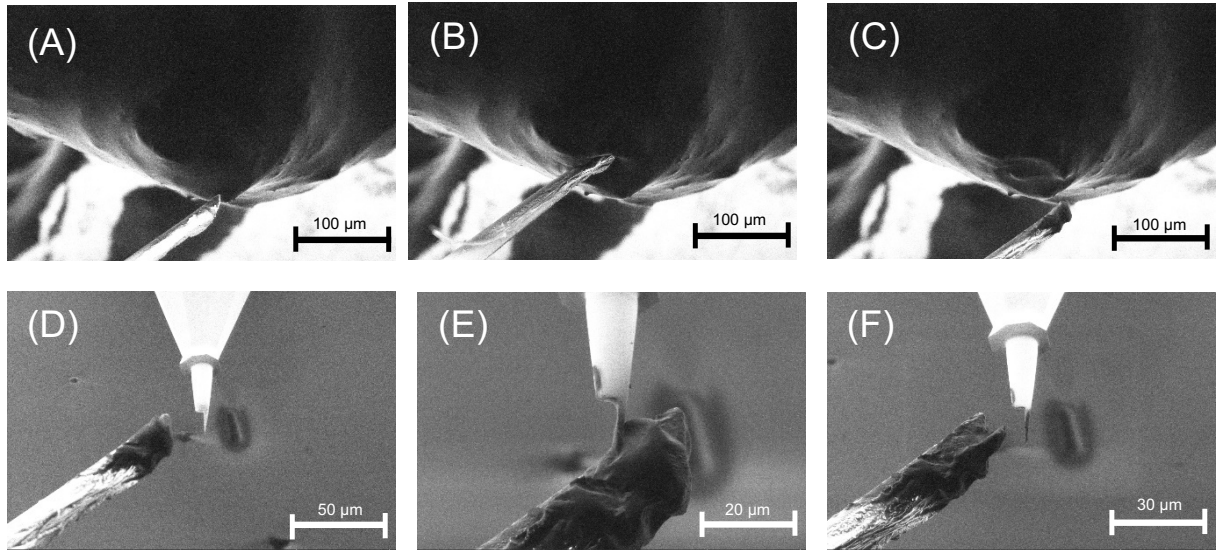

**Fig. 2:** SEM micrographs displaying (A)-(C) The micromanipulator being inserted and moving through a blob of SEMGlu™ at room temperature to pick up some of the glue. (D)-(F) the wiping process where the micromanipulator covered in SEMGlu™ is wiped across a precut Si microarray post until some of the glue remains on the post.

The above process does unfortunately require pre-preparation at room temperature which can add significant time to a cryogenic FIB session. Therefore, it was also investigated if this preparation could also be completed entirely once the instrument has cooled to cryogenic temperatures. Initially the viscosity of the SEMGlu™ at varying cooled temperatures was investigated. It was found that the glue remained viscous until ≈ 200K, meaning the glue was in a frozen state when operating at liquid nitrogen temperatures. It was found the glue could still be cured and a change in contrast could be noted when exposed to either electron or ion beams, meaning the glue can act as an adhesive even in its frozen form at cryogenic temperatures. Following these initial tests, the stage and micromanipulator were cooled to approximately 113K.

Inserting the micromanipulator into the SEMGlu™ is considerably more difficult at cryogenic temperatures due to this lack of viscosity. With sufficient pressure from the micromanipulator it is possible to fracture the top surface of the frozen glue. This process is easier to conduct using a pre-spread thin film rather than a large frozen blob of the glue. Further inserting the micromanipulator into the crack and moving the manipulator across the glue in a given direction it is possible to pick up some of the frozen glue. An example of this process can be seen in Figure 3(A-B). Oftentimes this method is not effective, meaning multiple attempts usually need to be made to try get some frozen glue to stay on the manipulator. It should be noted that due to the frozen nature of the glue it often is not possible to have a uniform layer of glue on the manipulator, making it difficult at times to make enough contact between the lift-out and the manipulator to get sufficient adhesion. Milling the glue to get a more uniform layer is of course possible but will result in the glue becoming cured, losing its adhesive capabilities. This method has been found to be particularly useful when more than one sample needs to be lifted-out in a given session. If the glue is picked up at room temperature and cured to lift something out at cryogenic temperatures, the glue on the manipulator is no longer useful. Picking up more glue while the FIB is cold allows more samples to be picked up easily.

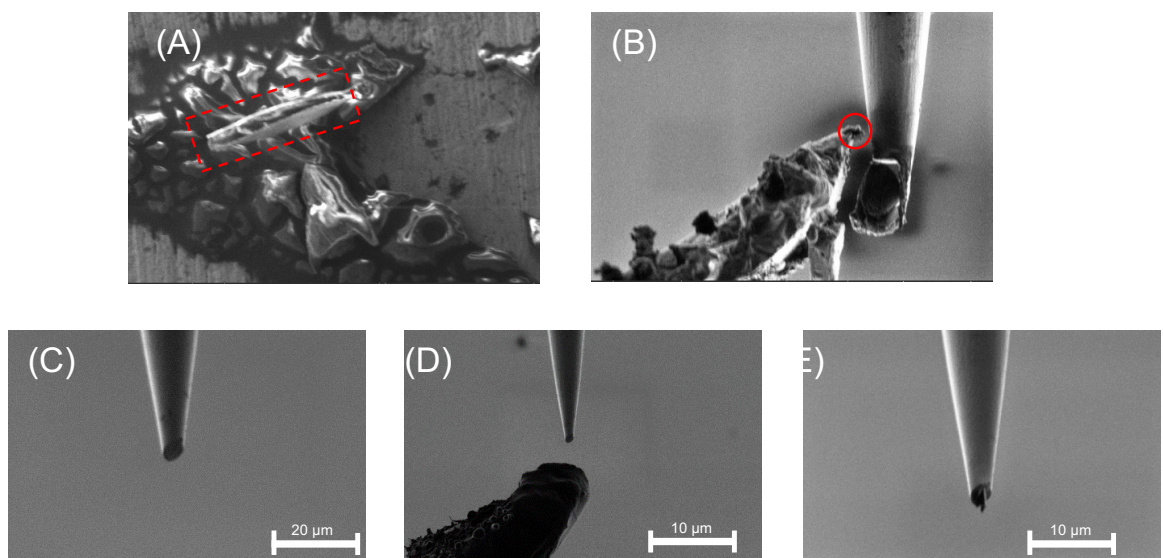

**Fig. 3:** SEM micrographs displaying (A) Fractured top surface of frozen SEMGlu™ at cryogenic temperatures. The micromanipulator has been placed in contact with pressure applied to fracture the surface and was subsequently dragged through the glue. (B) Frozen glue on micromanipulator picked up at cryogenic temperatures. (C)-(E) Wiping process at cryogenic temperatures and randomly orientated frozen glue wiped onto post is evident in (E).

Wiping the glue from the micromanipulator onto a Si post is a much more difficult task. While possible and shown here it is not recommended due to the random uniformity of a solid piece of glue attaching to the silicon post from the manipulator. When trying to attach a lift out to this solid piece of glue it can be difficult to fully cure the glue due to a lack of clear line of sight from either column. An example of these difficulties is evident in Figure 3(C-E). Moving two solid surfaces against one another within the FIB is also not recommended due to any potential damage that maybe occur to the Si post, or even the micromanipulator itself. While tedious and time consuming the authors recommend as much pre-preparation as possible at room temperature including covering the Si posts and the micromanipulator in glue. However, it is still possible to do both at cryogenic temperatures.

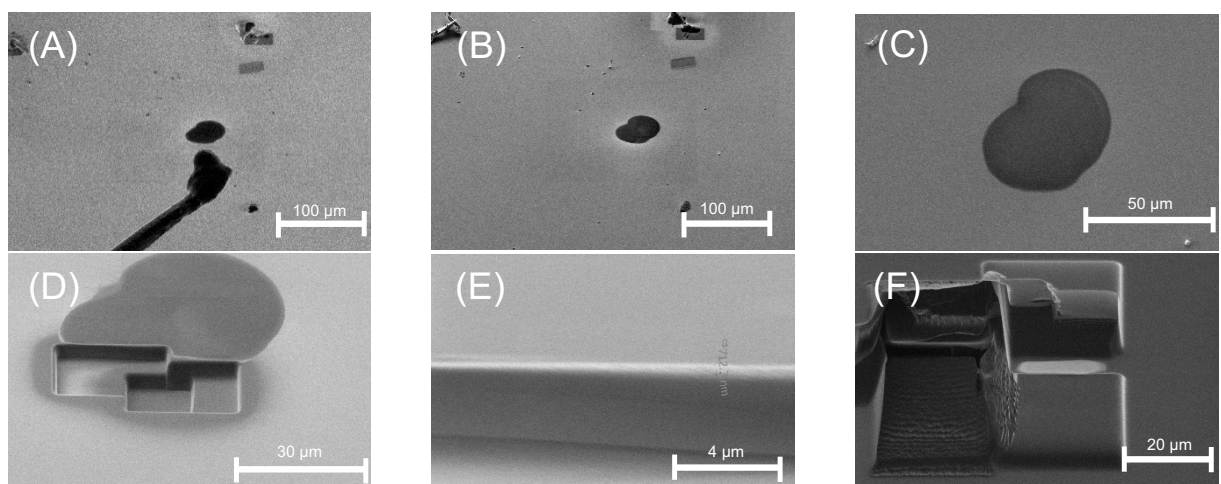

**Fig. 4:** SEM micrographs displaying (A)-(B) Thin layer of glue wiped on ROI at room temperature, (C) wiped glue cured using electron beam (10kV, 3.2nA), (D) glue milled at cryogenic temperatures to determine ability to act as protective layer, (E) cross section of protective layer showing a thickness of ≈ 700-800nm and (F) lift out bar prepared at cryogenic temperatures, showing glue on top acting as protective layer.

While difficult to wipe glue at cryogenic temperatures, materials which remain frozen below the point that the glue freezes (≈ 200K), can take advantage of its viscosity up to that point. One such advantage is using the glue to form a protective layer in place of GIS injected Pt. Wiping viscous glue over the region of interest and curing with the electron beam (10kV, 3.2nA) can produce a thin layer of material that can act as protection during milling at both room temperature and cryogenic temperatures. It was found that thin layers (≈ 1 μm) could be reproducibly created using this method. A thin protective layer prepared at room temperature by wiping the manipulator across a region of interest, cured, and milled into a lift out bar at cryogenic temperatures can be seen in Figure 4.

**Cryogenic Lift-out and Mounting of Model Si Sample**

To first test the viability of a cryogenic lift out involving SEMGlu™ and pre-prepared posts the entire procedure was performed at room temperature. A bar of Si was successfully lifted out using the glue on the micromanipulator and mounted to a Si post covered in glue. The sample required no additional adhesion to be lifted out or mounted.

Operating on the basis of the process working at room temperature, the lift out procedure was investigated at cryogenic temperatures. Following all pre-preparation at room temperature, the stage and micromanipulator were cooled to approximately 113K. A lift out bar approximately 30 $\mu m$ × 5 μm × 5 μm of Si was prepared on a Si microarray using typical milling patterns associated with APT sample preparation [56]. Both the undercut and lift out were performed at a stage tilt of 0° with respect to the electron beam. The pre-prepared frozen SEMGlu™ on the micromanipulator was brought in contact with the lift out bar. Due to the nature of the angles involved in the lift out process oftentimes the SEM beam could not access the exact point of contact between the glue and the lift out bar, meaning in most cases the ion beam (30kV, 30pA) was used to cure the glue. A change in contrast is evident when the glue has been cured. Often a good indication to determine if the micromanipulator is sufficiently stuck to the lift out bar is to change the current of the ion beam while viewing the sample using the electron beam. If the ion beam current is changed sufficiently the changing aperture will cause the micromanipulator to shake. If the manipulator is not sufficiently attached to the lift out it will "Shake it off" during this process. If this process is performed and the sample is not securely attached it can of course lead to the potential loss of a sample due to shaking the sample against the post. From here this will be referred to as the "Shake test". Once determined to be attached a portion approximately 5 $\mu m$ in length is milled free from the liftout bar and lifted out using the glue and manipulator. No movement or change in orientation of the sample was noted during this process and can be seen in Figure 5(A-B).

The sample attached to manipulator is then brought into contact with the frozen SEMGlu™ on a pre-prepared Si microarray post. The glue was again cured using the ion beam (30kV, 30pA) due to the angles involved, with a notable change in contrast observed as seen in Figure 5(D-E). The shake test was performed and the sample remained still indicating the sample was stuck to the post successfully. A portion of the lift out was then milled free from the micromanipulator and remained attached to the Si post without any further support, as in Figure 5(G).

To further test the viability of this cryogenic lift-out, the attached sample was prepared for atom probe tomography, as described in [17]. To that end, a notch underneath the sample was milled, and filled with tungsten from the micromanipulator using redeposition techniques. Once filled, using a series of annular milling patterns at lower and lower ion beam currents a needle shaped sample was created. This process can be seen in Figure 5(H-I). It was ensured as much SEMGlu™ as possible was removed from the sample during the annular milling process, once the sample had been filled in with an appropriate metal, in order to decrease the risk of thermal tail artefacts or increased carbon contamination, as described in [47]. While in usual practice a cryogenic sample would be transferred directly to the atom probe using a precooled VCTM, as described in [53], to maintain the sample in its frozen state. However for this demonstration cryogenic transfer was avoided to show this process is suitable for room temperature transfers also.

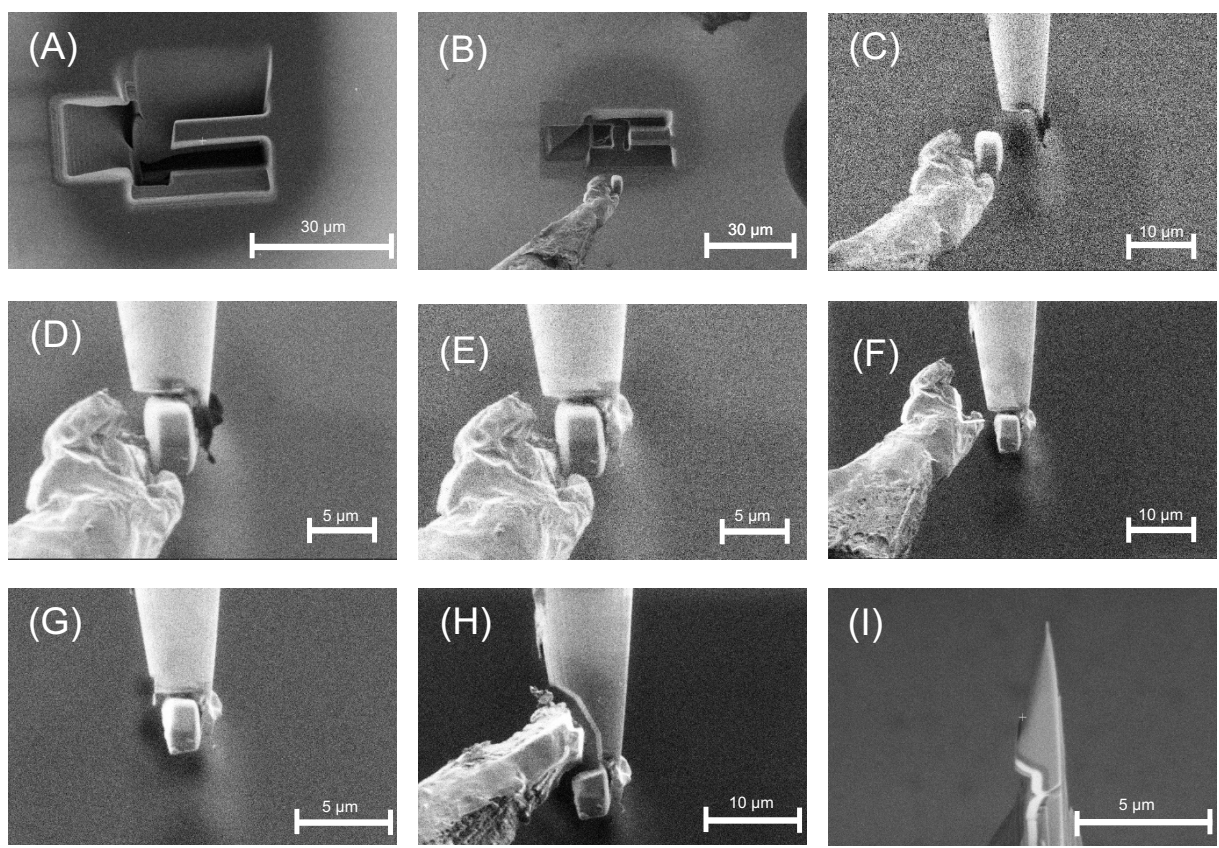

**Fig. 5:** SEM micrographs displaying (A) Lift out bar of Si prepared using ion beam at 52° at cryogenic temperatures. (B) Section of lift out bar lifted out using frozen glue on micromanipulator. (C)-(E) Lifted out section being brought in contact with frozen SEMGlu<sup>TM</sup> on precut preprepared Si microarray post. Note change in contrast of SEMGlu<sup>TM</sup> once cured between (D)-(E). (F) Micromanipulator cut free from lift out section. (G) Lift out section remains attached to Si post solely with SEMGlu<sup>TM</sup>. Position and orientation of lifted out section remains consistent throughout process. (H) Sample filled in with W from micromanipulator as described in . (I) Sample milled into needle shape appropriate for atom probe analysis.

The prepared sample initially underwent laser analysis (30 pJ, 140-200 kHz, 1 ion per 100 pulses on average, 50 K base temperature). The collected data showed comparable quality in comparison to a pre-sharpened microarray tip was analysed using the same conditions, as in Figure 6(A-B). Peak shapes and detected species remained consistent across both samples, while the SEMGlu<sup>TM</sup> prepared sample showed no indication of any carbon contamination due to the precaution of removing the glue during the annular milling process. In combination with the redeposited W, this indicated that the sample also did not present any visible "thermal tails" in the acquired data. This indicates that the sample was sufficiently thermally conductive. The same sample subsequently underwent voltage-pulsing analysis (20% pulse fraction, 200 kHz, 1 ion per 100 pulses on average, 50 K base temperature), reaching a standing voltage of 6 kV. The voltage-pulsing acquired data again showed comparable quality to the pre-sharpened microarray tip, as evident in Figure 6(C-D). The sample fractured at 6kV, potentially indicating the sample was not mechanically stable enough to reach higher voltages. This may be due to the geometry of this specific sample, with notches known to be stress concentrators. This could indicate that samples that require high voltages to achieve field evaporation may not be appropriate for this sample preparation workflow, but the sample thinning can be optimised to improve the upper threshold of voltages.

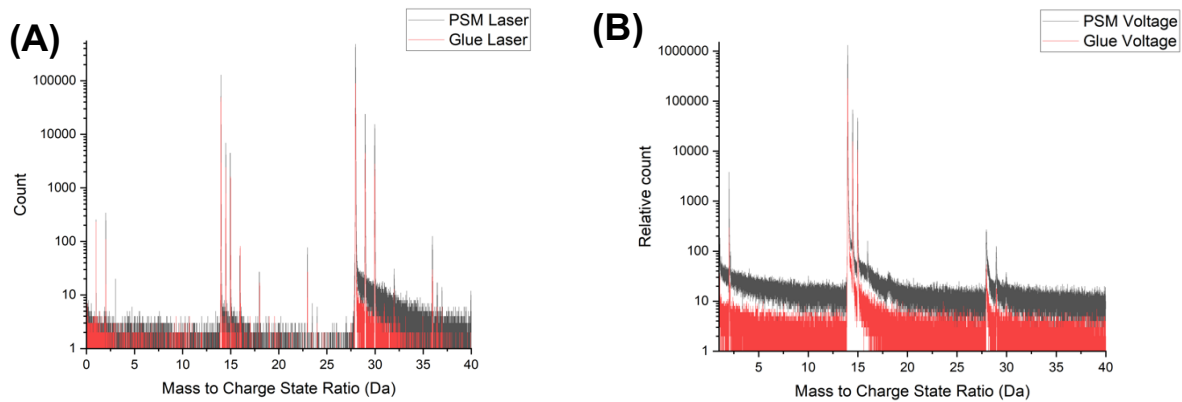

**Fig. 6:** (A) Mass spectra laser analysis of Si prepared using SEMGlu™ in red and of pre-sharpened microtip (PSM) in black. (B) HV pulsing mass spectra of Si prepared using SEMGlu™ in red and of pre-sharpened microtip (PSM) in black. Full dataset shown. Charge state ratio matching was not carried out.

Detection hit maps for both laser and voltage-pulsing analysis of both samples can also be seen in Figure 7. Again, the hit maps between both samples at both analysis modes look similar, with a distinct four-fold symmetry visible in all the hit maps. This symmetry can be attributed to the lift out being prepared from a [001]-oriented Si wafer. This symmetry is helpful for identifying the relative orientation of the sample on the post, with the main pole of the SEMGlu™ prepared sample being relatively central on the hit map. This indicates that the positioning and orientation of the lift out bar remained unchanged throughout the lift out and mounting process. This demonstrates reproducible and consistent positioning of the sample with respect to the post using this lift out and mounting method, which is especially important for samples with specific geometry requirements such as thin planar surfaces.

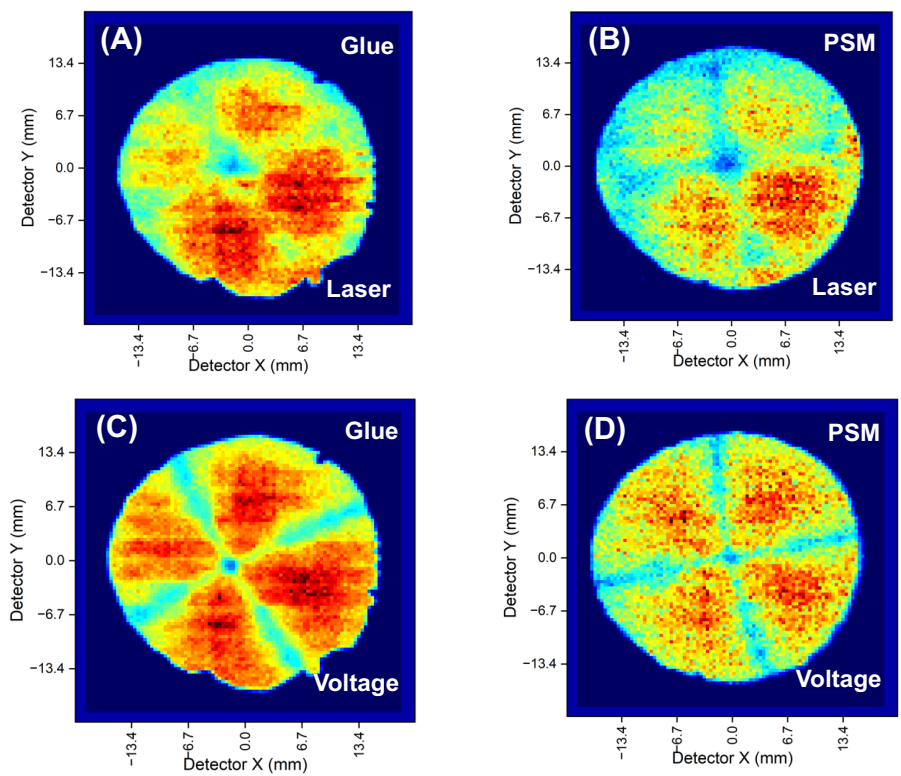

**Fig. 7:** Detector hit histogram from the analyses in laser pulsing mode of (A) SEMGlu™ prepared Si and (B) Pre-sharpened microtip, and in HV pulsing mode for (C) SEMGlu™ prepared Si and (D) Pre-sharpened microtip.

## Cryogenic liftout and mounting of application-based system – Correlative LCTEM and Cryogenic APT

While it has been shown that cryogenic lift outs are possible through the use of various redeposition techniques [3, 17], this has thus far only been demonstrated for samples where fixed geometries can be milled and lifted out. These samples can be undercut at specific angles to ensure there is sufficient contact between the Si post and sample to allow the redeposition to work effectively. Samples which can not necessarily be undercut or have an irregular shape are more difficult to attach to Si posts and using traditional redeposition methods oftentimes either fall off the Si post or move while being cut free from the manipulator, moving the region of interest due to insufficient overlapping area between the support structure and the sample.

A specific application that directly benefits from the development of a SEMGlu$^{TM}$ enabled cryogenic workflow is correlative LCTEM and cryogenic APT. The design of the LCTEM chips is such that the electrodes with grown electrochemical products are patterned on a thin silicon nitride membrane window (50 nm). This means when milling free one of these regions to produce a sample for APT analysis no undercut can be produced due to the thin layer of the membrane. This sample geometry, matched with the size of a typical electrode makes it exceptionally difficult to attach a lifted-out section to a Si post purely with redeposition techniques. Grown electrochemical products during the LCTEM experiments are often highly beam sensitive, making redeposition potentially destructive for these particular studies.

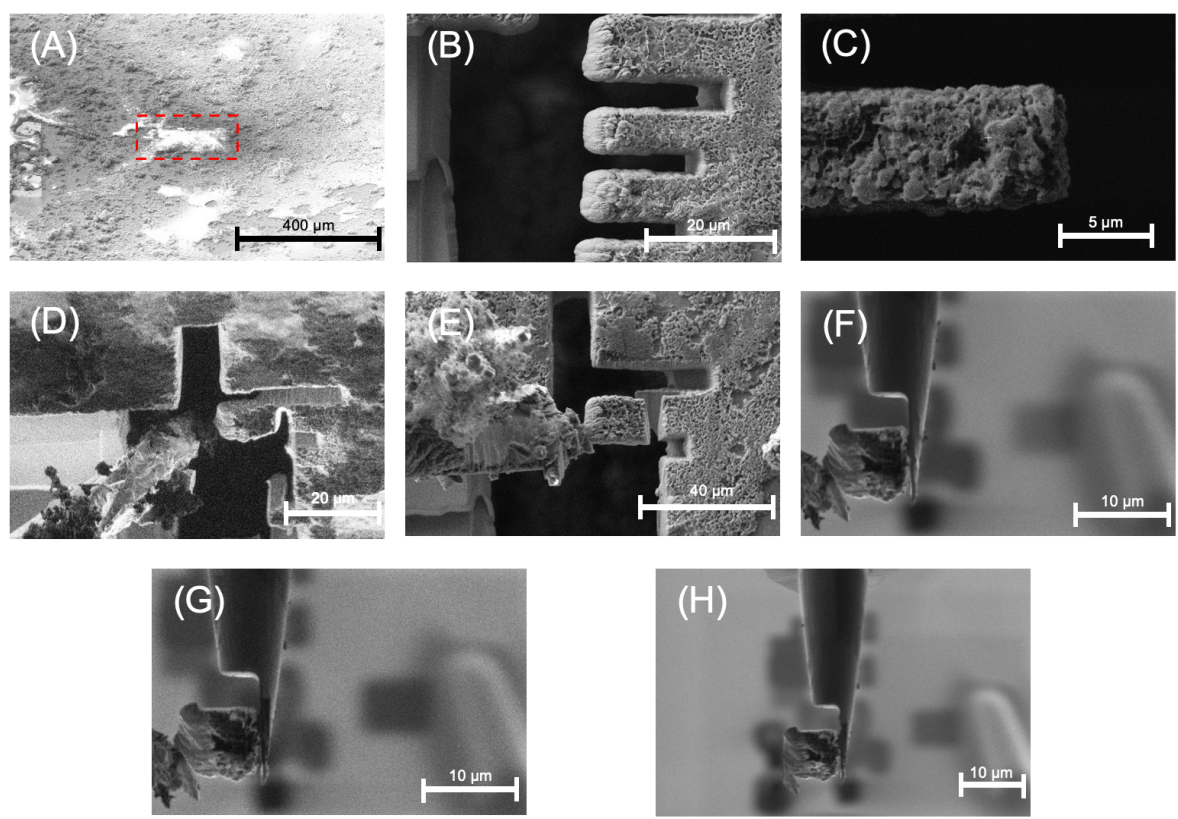

**Fig. 8:** Application based example of preparing samples for correlative LCTEM-Cryogenic APT using SEMGlu$^{TM}$ enabled cryogenic liftout and mounting. SEM micrographs displaying (A) Identifying ROI (i.e. frozen electrolyte-electrode interface). (B) Milling electrodes into shape to be lifted out. (C) Example of milled free electrode. Frozen electrolyte on top evident. (D)-(E) Lifting out electrode using frozen SEMGlu$^{TM}$. (F)-(G) Attaching frozen ROI to precut pre-prepared post using frozen SEMGlu$^{TM}$, note change in contrast evident following curing. (H) Frozen electrolyte-electrode interface stuck to preprepared post without further support and without losing orientation.

For this application precutting and precoating of the Si posts was performed at room temperature prior to cooling the instrument to 113K, while the SEMGlu<sup>TM</sup> on the micromanipulator was picked up at cryogenic temperatures as described above. Once cooled, the frozen liquid cell containing frozen Li electrolyte and grown electrochemical products was introduced to the cryo stage of the FIB from a precooled VCTM. The electrolyte-electrode interface was identified using the electron beam and then milled into sufficient shapes to both lift-out and maintain the interface using the ion beam (30kV, 1-4nA) at a 52° stage tilt. This process can be seen in Figure 8(A-C). Note the empty space below the milled free electrodes, indicating the unusual sample geometry and lack of potential undercut. The micromanipulator with frozen SEMGlu<sup>TM</sup> was brought into contact with the frozen electrolyte-electrode interface and cured using the ion beam (30kV, 30pA). A shake test was performed, followed by cutting the sample free and lifting it out. Note the maintained geometry of the sample in Figure 8(E). The frozen interface on micromanipulator was then brought in contact with the prepared Si post covered in frozen glue. The glue was again cured using the ion beam (30kV, 30pA) due to the angles involved. Note the evident change in contrast seen once cured, Figure 8(F-G). Following a shake test, the sample was milled free from the micromanipulator and remained stuck and in place attached to the Si post without further support. The sample orientation and position also remained consistent throughout the entire process. This sample preparation process was repeated numerous times for correlative LCTEM-Cryogenic APT sample preparation, producing reproducible success.

## Conclusions and Outlook

In conclusion, we have successfully demonstrated a GIS-free SEMGlu<sup>TM</sup> enabled methodology for atom probe specimen lift out and mounting at cryogenic temperatures. This process has been demonstrated for a model Si specimen where relative orientation and positioning, as well as thermal and mechanical stability has been shown to be comparable with a Si reference material from pre-sharpened microtips. The methodology was applied to an application-based system involving the preparation of samples for correlative LCTEM electrochemistry and cryogenic APT. This methodology greatly benefitted this application which was previously difficult to lift out and mount using pre-established methods due to the unique geometry presented by the sample of interest as well as the beam sensitive nature of the products within the sample. It is envisioned that there is a non-exhaustive list of specific applications which would directly benefit from this methodology, such as various particle-based battery systems, beam sensitive materials such as many biological samples, samples that mill easily such as various frozen liquids, systems with unique geometries such as many in-situ Si wafer chips, various porous systems, etc. While SEMGlu<sup>TM</sup> was not inherently designed for applications involving cryogenic temperatures, the curability of the glue at these temperatures, as shown here, opens up the field of using adhesives for these types of cryogenic applications. Potentially there could be a similar adhesive with better viscosity at liquid nitrogen temperatures that remains unknown to the community that could have the potential to make glue enabled GIS free cryogenic lift outs and mountings a more reproducible and accessible process.

## Acknowledgements

N.M. and M.C. acknowledge funding from the EPSRC InFUSE grant EP/V038044/1. This work was made possible by the EPSRC Cryo-Enabled Multi-microscopy for Nanoscale Analysis in the Engineering and Physical Sciences EP/V007661/1. M.C. acknowledges funding from Royal Society Tata University Research Fellowship (URF\R1\201318) and Royal Society Enhancement Award RF\ERE\210200EM1.

## Contributions

J.O.D. conceived the idea of using SEMglu<sup>TM</sup> to prepare samples under cryogenic conditions with a flat surface, N. M., J.O.D., M.C. conducted the SEM and FIB experiments, N.M., and M.C. did the in-situ TEM experiments, N.M. and J.O.D. analysed the APT specimens and processed the data. All authors discussed the results and contributed to the final version of the manuscript. J.O.D. and M.C. contributed equally as senior last corresponding authors.

## Corresponding authors

James Douglas, j.douglas@imperial.ac.uk

Michele Conroy, mconroy@imperial.ac.uk

## Conflict of Interest

The authors declare that they have no known competing financial interests or personal relationships that could have appeared to influence the work reported in this paper.